\documentclass[journal, a4paper, 12pt]{IEEEtran} %
\IEEEoverridecommandlockouts
\usepackage{cite}
\usepackage{amsmath,amssymb,amsfonts}
\usepackage{minipage-marginpar}
\usepackage{enumitem}
\usepackage{graphicx}
\usepackage{textcomp}
\usepackage{xcolor}
\usepackage{hyperref}       
\usepackage{url}            
\usepackage{booktabs}       
\usepackage{amsfonts}       
\usepackage{mathtools}
\usepackage{float}
\usepackage{dblfloatfix}    
\usepackage{multirow}
\usepackage{algpseudocode}
\usepackage[ruled,vlined,linesnumbered]{algorithm2e}

\usepackage{amsthm}
\usepackage{etoolbox}

\AfterEndEnvironment{theorem}{\noindent\ignorespaces}

\AfterEndEnvironment{corollary}{\noindent\ignorespaces}

\AfterEndEnvironment{lemma}{\noindent\ignorespaces}

\AfterEndEnvironment{definition}{\noindent\ignorespaces}

\AfterEndEnvironment{example}{\noindent\ignorespaces}
\usepackage{amssymb}
\usepackage{amsmath}

\usepackage{xfrac}
\usepackage{array}
\newcolumntype{x}[1]{>{\centering\arraybackslash\hspace{0pt}}p{#1}}

\def\BibTeX{{\rm B\kern-.05em{\sc i\kern-.025em b}\kern-.08em
	T\kern-.1667em\lower.7ex\hbox{E}\kern-.125emX}}
\begin{document}

%
\title{Personal Devices for Contact Tracing: Smartphones and Wearables to Fight Covid-19}

\author{Pai Chet Ng,~\IEEEmembership{Student Member,~IEEE,}
	Petros Spachos,~\IEEEmembership{Senior Member,~IEEE,} \\
	Stefano Gregori,~\IEEEmembership{Senior Member,~IEEE,} and 
	Konstantinos N. Plataniotis,~\IEEEmembership{Fellow,~IEEE}
}

\maketitle
\begin{abstract}
	Digital contact tracing has emerged as a viable tool supplementing manual contact tracing.
	To date, more than 100 contact tracing applications have been published to slow down the spread of highly contagious Covid-19.
	Despite subtle variabilities among these applications, all of them achieve contact tracing by manipulating the following three components: a) use a personal device to identify the user while designing a secure protocol to anonymize the user's identity; b) leverage networking technologies to analyze and store the data; c) exploit rich sensing features on the user device to detect the interaction among users and thus estimate the exposure risk.
	This paper reviews the current digital contact tracing based on these three components. We focus on two personal devices that are intimate to the user: smartphones and wearables. We discuss the centralized and decentralized networking approaches that use to facilitate the data flow. Lastly, we investigate the sensing feature available on smartphones and wearables to detect the proximity between any two users and present experiments comparing the proximity sensing performance between these two personal devices.
\end{abstract}

\IEEEpeerreviewmaketitle

\section{Introduction}
\label{sec:intro}
\noindent
\IEEEPARstart{2}{020} will be a long-remembered year to many as the outbreak of a global pandemic has severely affected millions of lives.
Besides imposing restriction measures, from small scale (city-wide) to large scale (country-wide), many countries started to exploit digital contact tracing to supplement the laborious contact tracing performed manually by human investigators~\cite{ferretti2020quantifying, eames2003contact}.
These digital contact tracing applications exploit rich sensing features from a device that is either carried or worn by users to detect the proximity between users while anonymizing users' identities. Networking technologies are leveraged to facilitate the data flow while identifying potential exposures

\begin{figure}[t!]
	\centering
	\includegraphics[width=1\columnwidth]{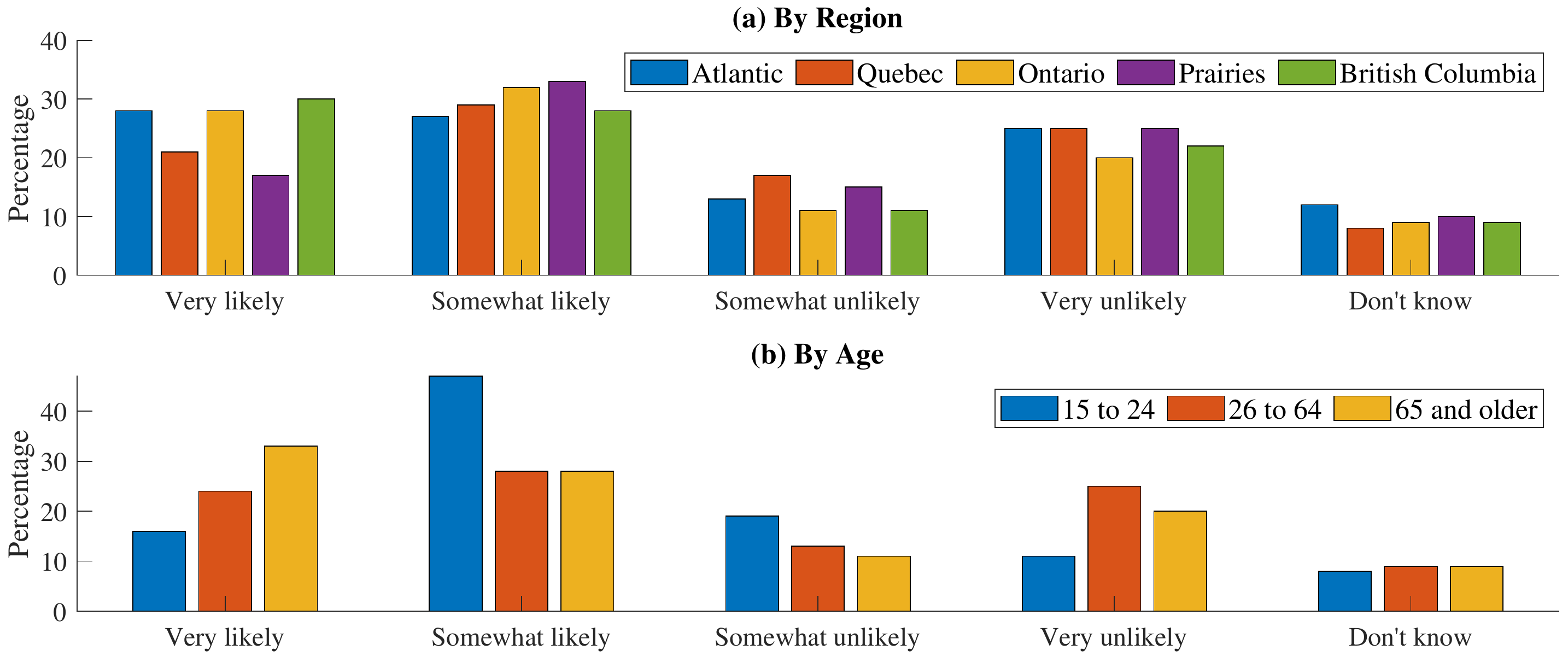}
	\vspace*{-0.3cm}
	\caption{The willingness of Canadians to use the contact tracing application (a) by region, and (b) by age.}
	\label{fig:digitalCT}
\end{figure}

\begin{table*}
	\caption{Contact Tracing Applications launched by each Country.}
	\label{tbl:ctApp}
	\centering
	\includegraphics[width=0.75\linewidth]{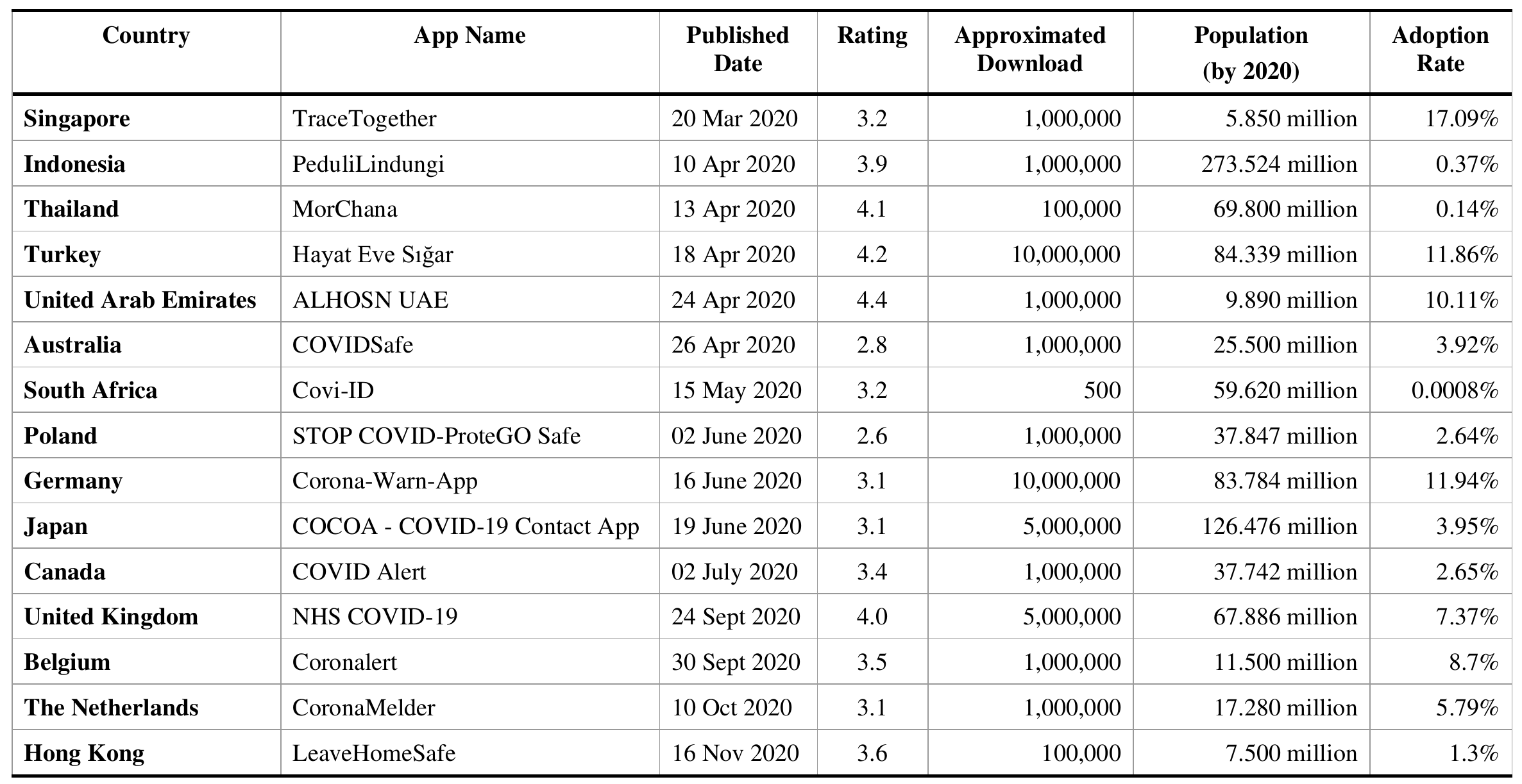}
\end{table*}

A typical digital contact tracing can be described by three components: a) user devices, b) networking technologies, and c) sensing features.
The user device should be a device personal to the user and carried by the user most of the time. 
In other words, we need a personal device that is capable of representing the user uniquely, besides providing rich sensing features and networking capabilities.
Two prominent user devices that satisfy the above criteria are smartphones and wearables.
Since these devices always contain sensitive data about users, the privacy issue has been the main concern when implementing contact tracing solutions onto either of these two devices.
Many different privacy-preserving protocols have been designed to anonymize the user's identity while using the personal device to identify the high-risk user who may have been exposed to the virus due to their proximity to the infected patient.

While myriad contact tracing applications have been published in Google Play Store and Apple App Store, the public acceptance of digital contact tracing is quite low, resulting in the low installing rate of these App.
Fig.~\ref{fig:digitalCT} shows the willingness of Canadians to adopt the contact tracing application according to the Canadian Perspectives Survey conducted in June 2020 \cite{StatCan}.
Even though many people in Ontario and other regions indicating their willingness to use the application, most of them are only somewhat likely to install it. The young working adult group (aged between 26 and 64) tends not to support the contact tracing initiative due to privacy and performance concerns. For example, they worry about installing a surveillance tool into their device, and they doubt the reliability of the application in measuring the proximity.

In view of this, this paper studies the current digital contact tracing from the perspective of the three components described above. 
The objective of this paper is to discuss the privacy and performance issues, which are two main concerns that deter users from installing the application.
Experiments are presented to verify the performance issue comparing the contact tracing solution implemented into smartphones and smartwatches.
Lastly, a few possible research directions are suggested to improve the digital contact tracing solution so that it can become a mature tool to combat not only the Covid-19 but also any highly contagious diseases that we may have in the future.

\section{Personal Devices for Contact Tracing}
\label{sec:ct}
Manual contact tracing, conducted by professional investigators, is time-consuming and not effective considering the short-term memory of human being~\cite{ferretti2020quantifying}. 
For example, can you recall who you have talked to when you were doing your grocery shopping, and would you be able to get their contact information? 
Rather than relying on our short-term memory, digital contact tracing relies on users' devices to record daily interactions and store this information into digital-based storage, which can be retrieved automatically once a user is tested positive.
As discussed, this device should be a personal device that can represent the user uniquely wherever the user goes.

\subsection{Digital Contact Tracing with Smartphones}
To date, more than 50 countries have launched their contact tracing applications, and all of them are using the smartphone as the personal device representing the user.
Using the approximated downloading data from the Google Play Store, we summarized a few contact tracing applications, as illustrated in Table~\ref{tbl:ctApp}.
Since the download action in Google Play Store is equivalent to an install action (i.e., Google Play Store will download and install the App for the user when the user clicked the ``get" button), we can use the downloading data to compute the adoption rate.
Even though all of the applications have ratings of at least 2.5 (Google and Apple allow users to rate the App according to the rating scale from 1 to 5, in which 1 indicates the least satisfaction and 5 the most satisfaction), users generally commented that they have negative experiences with the applications.
Hence, the rating itself might not necessarily reflect the users' satisfaction with the application.
Instead, we can examine the application's adoption rate since it is published. 
Note that the computed adoption rate here is solely based on the downloading data from Google play store as Apple provides no such information.
Even though we only got the adoption rate based on the Google play store, it should give us a big picture of the general adoption rate as a whole since Android smartphones have the majority of users in the world and the App downloading trend between iOS and Android devices exhibit a linear correlation.

Table~\ref{tbl:ctApp} shows that the adoption rate is pretty low even for the application launched in the first half of 2020 (e.g., TraceTogether by Singapore, PeduliLindungi by Indonesia, MorChana by Thailand, COVIDSafe by Australia, Hayat Eve Sigar by Turkey, and Covi-ID by South Africa). 
Comparing to Asia Pacific countries, most of the European countries only launched contact tracing applications in the second half of 2020. 
However, some of these applications have a higher adoption rate even though they are launched at a much later date.
This is mostly due to the privacy-preserving feature implemented into the application that gave users a certain assurance about their privacy. 
Even though these applications have fewer privacy issues, they suffer from performance issues, which causes users to uninstall the applications after an initial attempt.

\begin{table}
	\caption{Wearable-based Contact Tracing}
	\label{tbl:ctWearable}
	\centering
	\includegraphics[width=1\columnwidth]{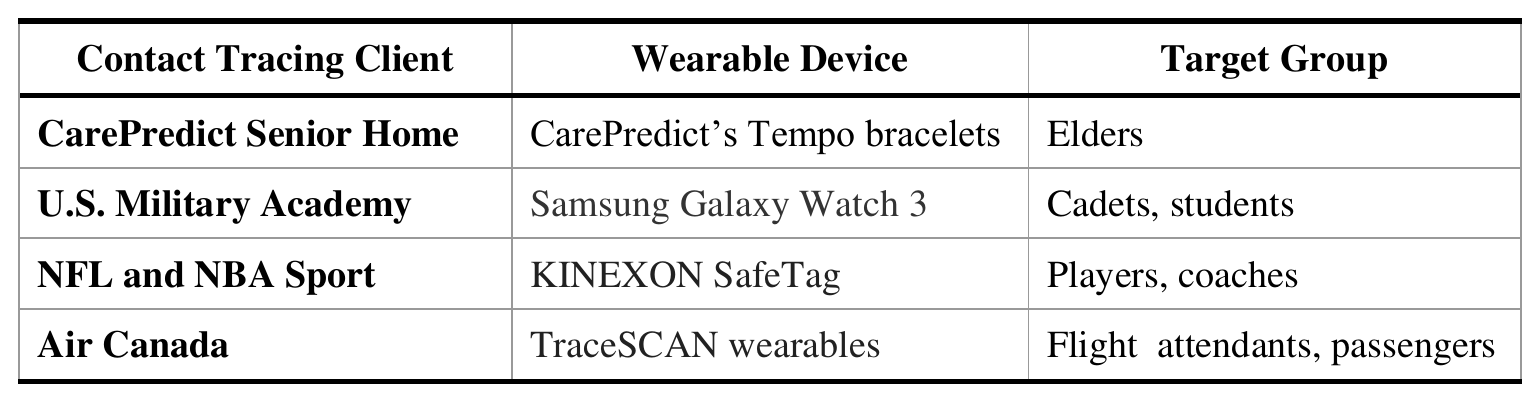}
\end{table}
\begin{figure*}[t!]
	\centering
	\includegraphics[width=.9\linewidth]{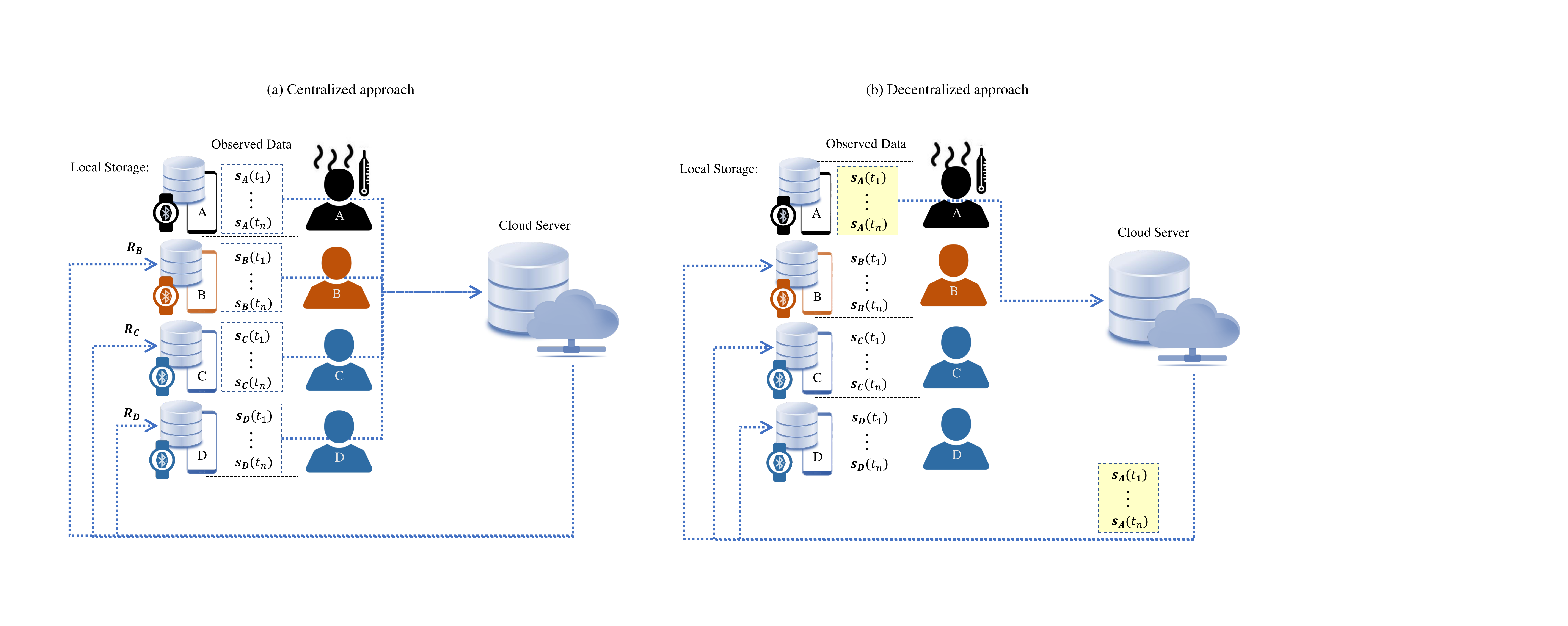}
	\caption{Networking technologies in facilitating the data flow: a) the centralized approach uses the cloud to store all the interaction data from all the users, wheres b) the decentralized approach stores the interaction data within the local storage of each individual device, and the cloud is used to broadcast the data from the infected patient to a group of devices that have the App installed.}
	\label{fig:networking2}
\end{figure*}
\subsection{Digital Contact Tracing with Wearables}
Comparing to large-scale implementation of contact tracing applications with smartphones, wearable-based solutions focus on a much smaller scale, and most of the time, they mainly trace the interaction of a specific group of users in a specific location.
Besides utilizing physiological signals~\cite{mishra2020pre}, or activity tracker~\cite{quer2020wearable}  from the wearable device to detect infection symptoms, many workplaces start to exploit the wearable solution for contact tracing~\cite{9270632}.
While almost everybody owns at least a smartphone, the penetration of smartwatches is relatively low, i.e., only 4 out of 10 people have a smartwatch (in the USA), or about 2 out of 10 people (in developing countries). 
Hence, the impact of releasing wearable-based contact tracing applications in the application store would be limited since not many people owning a smartwatch.
Rather, most of the current wearable solutions are custom-made and target dense indoor environments like hospitals, shopping malls, or workplaces.
For example, \cite{9278701} uses a wearable tag worn on users' wrist to track and monitor the health-care workers working in a hospital or a health-care center, which are at high risk of exposure to virus infection.
Ng et. al. presents a wearable solution based on a commercial off-the-shelf smartwatch that can be adopted by any industry towards workplace reopening \cite{ng2020epidemic}.

While the smartphone-based solution is targeting each individual, the wearable-based solution is targeting industrial sectors, expecting them to purchase the wearable device installed with the contact tracing application in bulk and distribute the wearable device to their workers and customers.
Table~\ref{tbl:ctWearable} lists a few industries that have adopted the wearable-based contact tracing solution to protect elders in the nursing home, cadets in the academy campus, players in the sports center, and passengers in the airline.
In fact, many active companies in IoT and wireless solutions, including Estimote, Radiant, etc., have launched their wearable-based contact tracing solutions. 
However, the industrial adoption rate is relatively small compared to the multitude of wearable solutions on the market today.

\section{Networking Technologies for \\Contact Tracing}
The second component of contact tracing relies on networking technologies to facilitate the data flow between the user device and the cloud server, either in a centralized or decentralized manner.

\subsection{Centralized Approach}
As shown in Fig.~\ref{fig:networking2}(a), the centralized approach uses a centralized server to store all the data uploaded from users' devices. 
To preserve privacy, the data are mostly encrypted before uploading to the central server for storage. 
Once a user is diagnosed with the disease, the server will perform the matching computation, and only send the alert to a list of users that have been identified as high-risk users who are likely to contract the virus.

An example of contact tracing that uses the centralized approach is the EasyBand~\cite{9085930}, in which the data from each wearable device is uploaded to the centralized server for further processing.
Besides storing the encrypted data uploaded by the user devices, the centralized server also performs the matching computation to identify a list of high-risk users.
The Pan European Privacy-Preserving Proximity Tracing (PEPP-PT)~\cite{PEPP} is also a centralized approach focusing on the proximity sensing between any two smartphones. PEPP-PT only requires the smartphone to upload the data when it comes into close proximity with another smartphone and stays close for a certain duration. 
Most of these centralized approaches guarantee users that only encrypted data will be collected, and there is no information regarding the user's location and sensitive information.

\subsection{Decentralized Approach}
The decentralized approach allows each device to store the daily interaction data into their device's local storage, as illustrated in Fig.~\ref{fig:networking2}(b).
All these data will remain in the local storage for 14 to 21 days depending on the design of the application, any expired data (i.e., more than the defined timespan) will be erased from the local storage.
Once a user is diagnosed with Covid-19, he/she can upload his daily interaction data to the cloud server, which is responsible to distribute uploaded data to all the user devices that have registered with the cloud server.
The data uploading processing is subject to the user's consent. If user declined to upload the data, the data would not be uploaded to the cloud.
Once received the data, the cloud server will broadcast the data to all the registered devices without having each device querying the cloud periodically.

Many smartphone-based contact tracing applications developed by European countries (e.g., Switzerland, Finland, etc.) mostly apply the decentralized privacy-preserving proximity tracing protocol, known as DP-3T~\cite{troncoso2020decentralized}.
Inspired by DP-3T, Google and Apple jointly developed a decentralized exposure notification framework to facilitate the development of contact tracing applications.
To date, many countries, including Canada, UK, etc., have adopted the Google and Apple framework to provide a decentralized-based contact tracing.
According to \cite{CoApps}, people are more willing to install the contact tracing application when they are assured that the application is developed via the decentralized approach, in which all the data will remain the local storage and will be erased after a certain timespan.

\section{Proximity Sensing \\for Contact Tracing}
Contact tracing records the daily interaction data containing the proximity information, as well as how long the two users remain in close proximity.
Various sensing features available on smartphones and wearable devices can be used for proximity sensing, including the radio signal, the magnetic signal, the acoustic signal, etc.
This paper mainly focuses on the radio signal from Bluetooth Low Energy (BLE) transmission considering the vast majority of applications listed in Table~\ref{tbl:ctApp} use BLE as the main sensing feature for proximity detection.
Hence, it is crucial to understand the performance issue of using BLE signals for proximity detection.

\subsection{Fluctuation of Received Signal Strength}
Proximity sensing via BLE signals refers to the measurement of smartphones’ Received Signal Strengths (RSS) to estimate their separation.
The issue of using RSS for proximity sensing is that RSS is susceptible to environmental dynamics, in which the measurement value fluctuates even though two devices are placed stationary during the measurement process~\cite{7174982, 6856188}.
Besides the environmental variations, device diversity is another problem affecting the final measurement value.
One way to mitigate the problem is to calibrate the model before applying the model to estimate the proximity given the input RSS value. 
However, such a calibration approach is only applicable if we have obtained sufficient data describing the device characteristics.

\begin{figure}[t!]
	\centering
	\includegraphics[width=0.9\columnwidth]{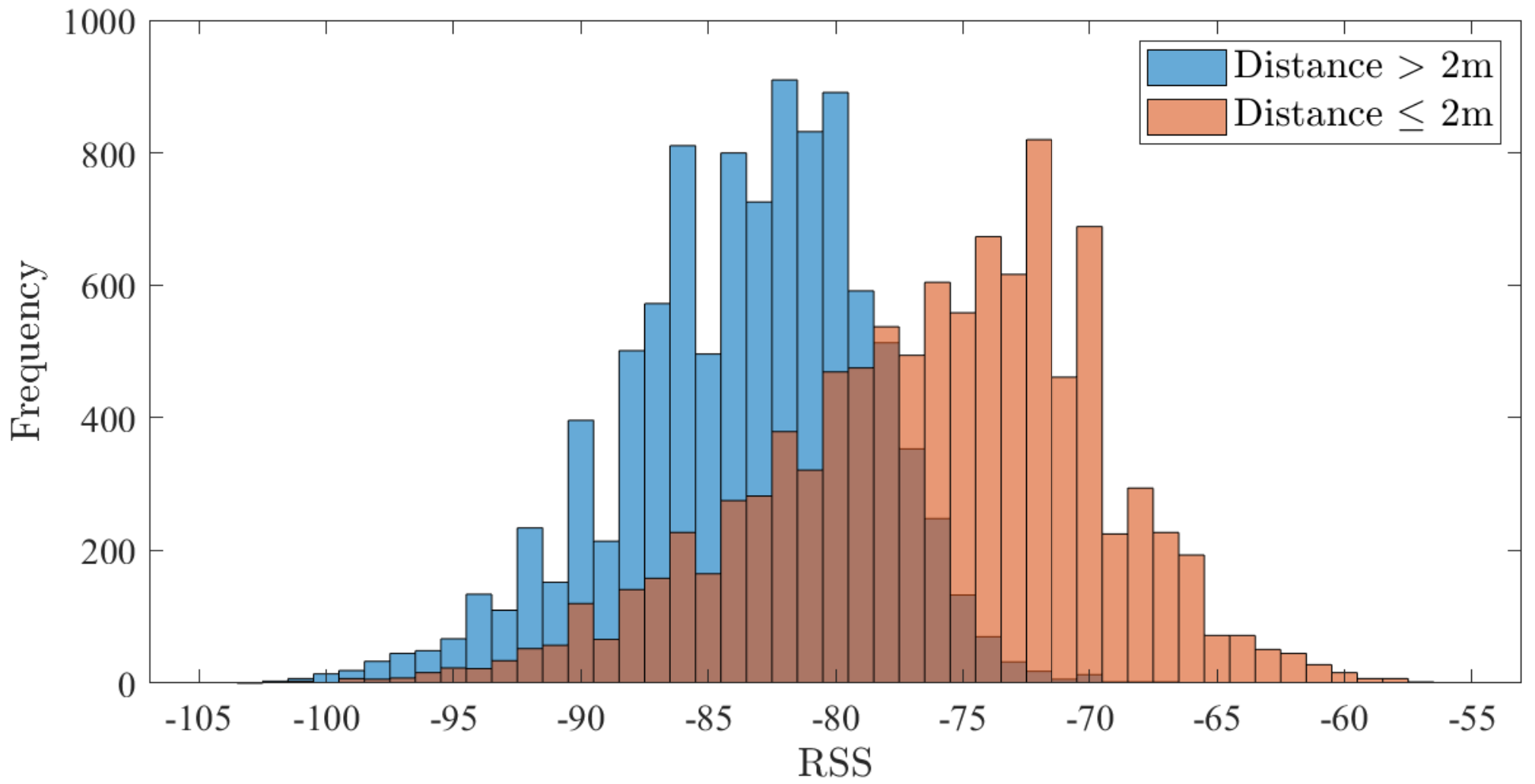}
	\caption{RSS distributions for two types of proximity: far (blue color bars) and close (orange color bars).}
	\label{fig:cutDownRSSdistribution}
\end{figure}
\subsection{Data-driven Approach} 
Many works have exploited data-driven approaches to improve proximity sensing with BLE signals. 
In particular, BLE data are collected to train a classifier that can classify the proximity between any two persons into the following two categories: close or far. 
The cutoff distance that separates the close and far is always based on the social distancing rule suggested by the government. 
In Canada, any two persons who stand less than 2~m from each other are considered in close proximity. 
According to Fig.~\ref{fig:cutDownRSSdistribution}, such a direct cutoff distance may produce a lot of misclassification.
Suppose that -80~dBm is the cutoff threshold for the distance greater than 2~m.
Fig.~\ref{fig:cutDownRSSdistribution} shows that some devices that are very close to each other (i.e., having a distance less than 2~m) can record an RSS value less than -80~dBm.
This means that the application will consider users far away from each other, but they are, in fact, close to each other, resulting in false negatives.
Rather than relying on the raw RSS measurements, we can extract more input features, such as the time indicating the interaction duration, the device model, etc., to train a more robust classification.

\section{BLE Dataset from \\Smartphones and Smartwatches}
\label{sec:imp}
The lack of comprehensive datasets based on BLE signals is the major drawback preventing the further development of machine learning methods for proximity sensing. 
This paper describes two BLE datasets (one is based on smartphones, and another is based on smartwatches) collected from our previous works \cite{9373368}, and then discusses the experimental results based on a supervised classification model trained using our datasets.
These two datasets are made publicly available to encourage further research, not limited to contact tracing applications, but any IoT applications that deal with BLE signals.

\subsection{Smartphones Dataset}
We collected a large-scale BLE dataset by having smartphones placed in different positions. 
Specifically, two volunteers were required to stand at a distance from each other while holding the smartphones in their hands. The ground truth distance is measured using a measuring tape. 
The application starts the scan when the user presses the scan button. The scan continues until the user press the button again.
We repeated the same data collection procedures for distance from 0.2~m to 2.0~m (with a 0.2~m increment each step), and 3~m to 5~m (with a 1~m increment each step).
A total of 13 distance points where the measurement is conducted.
For each distance, the smartphone was configured to run for at least 60~s.
During the 60~s measurement, the user is not subject to any restrictions, they can check their message, listen to music, talk to others, etc.

We repeated the data collection by asking the volunteers to put the smartphone in different positions. 
Besides ``Hand-to-Hand (HH)", we also collected the data for another five different position combinations, including ``Hand-to-Pocket (HP)", ``Hand-to-Backpack (HB)", ``Pocket-to-Backpack (PB):", ``Pocket-to-Pocket (PP)", and ``Backpack-to-Backpack (BB)". 
In total, we have collected 123,718 data points. 
We have made our collected dataset publicly available at \url{https://github.com/pc-ng/rss\_HumanHuman}.

\subsection{Smartwatches Dataset}
We developed a smartwatch App and installed it into Fossil Sport, which is powered by Google's Wear OS, to broadcast and collect the BLE data.
We performed the experiment by asking two volunteers to stand at a certain distance from each other, from 0.5~m up to 5~m. 
The data collection was performed in indoor environments with a lot of interference and at a different time with uncontrolled environmental settings so that the dataset can capture the signal distortion subject to the environmental dynamics. 
During the data collection, volunteer A was requested to wear the smartwatch on her left hand, and volunteer B on her right hand (i.e., left to right (LR)). After that, we repeated the same procedures with right hand to left hand (RL), left hand to left hand (LL), and right hand to right hand (RR).

Since LR and RL constitute a direct view between two smartwatches and LL and RR constitute the crosswise view, we categorize these four hand-combinations into two groups: a) direct, and b) crosswise.
In total, we have collected 37,644 data points from all four combinations. 
We consolidated the data from RR and LL into a single dataset (i.e., the crosswise dataset) and then apply an 80\%-20\% splitting rule to split the data into training and testing sets.
Similarly, we applied the same splitting rule to the consolidated data from RL and LR (i.e., the direct dataset). 
The final training and testing data for both sets are shared openly in \url{https://github.com/pc-ng/rss_smartwatch}.

\begin{table}[t!]
	\caption{Comparison between smartphone and smartwatch approaches}
	\label{table:compare}
	\centering  
	\begin{tabular}{cx{3cm}rx{3cm}rx{3cm}}  
		\toprule
		\cmidrule{1-3}
		Approach     &  Combination & Accuracy  \\					
		\midrule
		Smartphone   	& hand-to-hand            & 85.82\%\\
		Smartphone      & hand-to-pocket          & 90.75\%\\
		Smartphone   	& hand-to-backpack        & 81.44\%\\
		Smartphone   	& pocket-to-backpack      & 87.51\%\\
		Smartphone   	& pocket-to-pocket        & 87.26\%\\
		Smartphone 		& backpack-to-backpack    & 90.85\%\\
		\midrule
		Smartwatch   	& direct                  & \textbf{94.16}\% \\
		Smartwatch   	& crosswise               & 90.59\% \\
		\bottomrule
	\end{tabular}
\end{table}
\subsection{Classification Performance}
We train a decision tree classification model given the training data from both datasets and then applying the trained model to the testing data to examine the classification performance. 
For our experiment, we defined the cut-off distance to 2 m as this is the distancing rule set out by the Canadian Government.
Our experimental results in Table~\ref{table:compare} show that the best classification result is achieved when the smartphone was held in a similar manner.
The accuracy is 90.85\% when both users put their smartphones inside the backpack, 90.75\% when one user held the smartphone in their hand and another user put the smartphone inside their pocket, and 85.82\% when both users hold their smartphones on their hands.
The performance with hand-to-hand is not good compared to the rest. This can be explained by the subtle hand's movement when holding the smartphone. For example, some people might hold their smartphones above their arms, some use two hands to text, some use one hand to open up the notification, etc.
On the other hand, the smartphone is subject to fewer variations when the smartphone is put inside the backpack or pocket. 
Compared to the smartphone, the smartwatch can have a better performance most of the time because the smartwatch is always worn on a human's wrist, and there are relatively less variations compared to the smartphone. 
The classification performance is boosted from an average 90\% with the smartphone approach to approximately 94\%.

\section{Conclusion}
\label{sec:conclusions}
Digital contact tracing can be a prominent solution to slow down the viral spread of the contagious virus if more people are willing to install the application. 
This paper discusses two personal devices (i.e., smartphones and smartwatches) that have been widely exploited for digital contact tracing.
Note that smartphone-based and wearable-based solutions are targeting different groups of users for different scenarios. 
While smartphone-based solutions are mostly focusing on our day-to-day life, wearable-based solutions are specifically designed to assist the workplace reopening. 
So far, these two solutions are available to the public independently without any integration. 
Future work can consider the integration of these two solutions so that to provide a continuous contact tracing covering a person's working (based on wearable solution) and daily life routines (based on smartphone solution).



\section*{Acknowledgment}
This project is funded by NSERC Alliance COVID-19 grant \# 552130-20.



%
%
%

\bibliographystyle{IEEEtran}
\bibliography{references}

%
\begin{IEEEbiography}[{\includegraphics[width=1in,height=1.25in,clip,keepaspectratio]{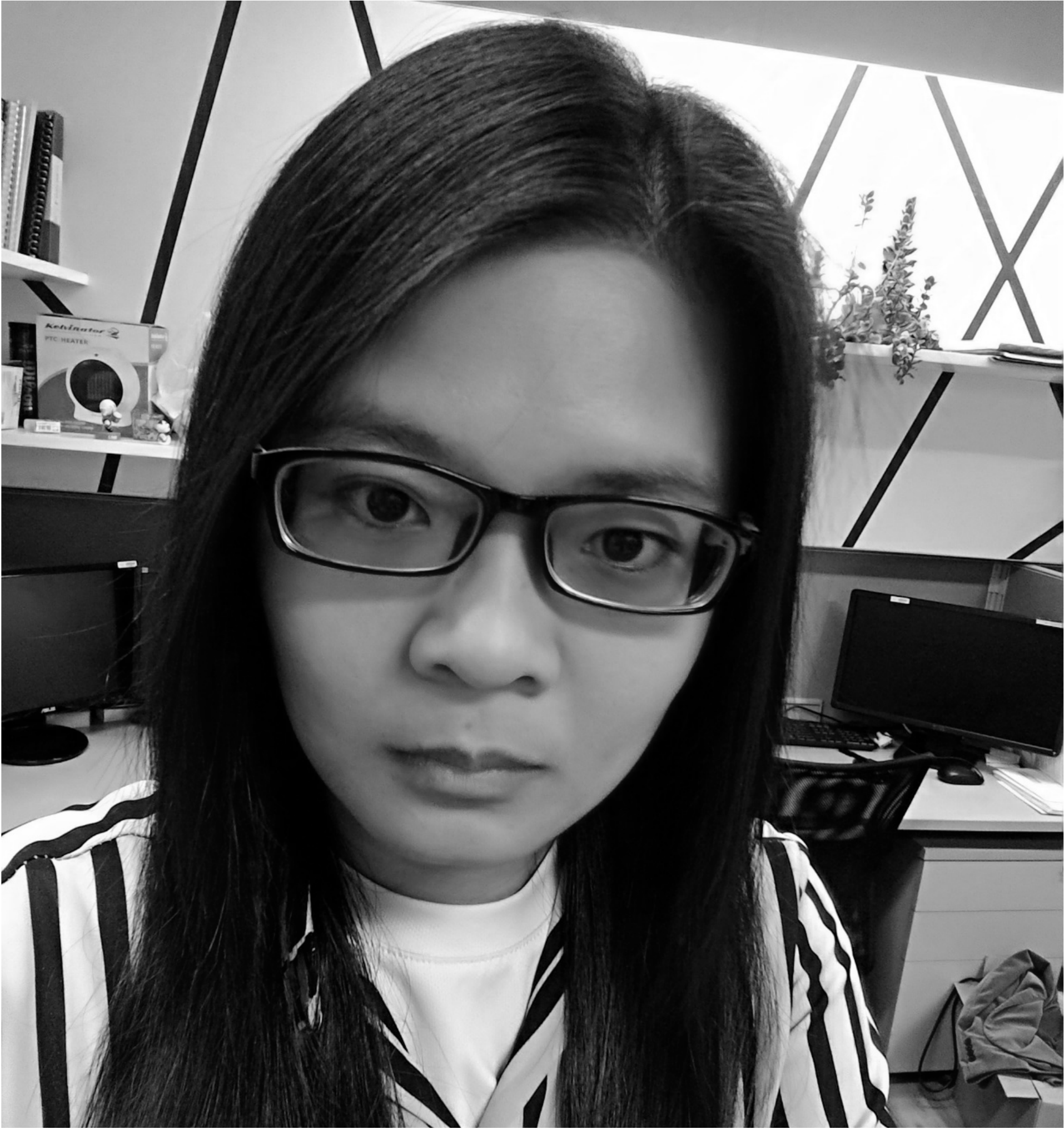}}]{Pai Chet Ng}
	is currently working toward the PhD degree in the Department of Electronic and Computer Engineering at the Hong Kong University of Science and Technology (HKUST). She received her BS degree in Telecommunication Engineering from Multimedia University, Malaysia. 
	She worked as research engineer prior to joining the HKUST-NIE Social Media Lab. Her research interests include RF signal processing, mobile and IoT analytics.
\end{IEEEbiography}
\vspace*{-0.5cm}

\begin{IEEEbiography}[{\includegraphics[width=1in,height=1.25in,clip,keepaspectratio]{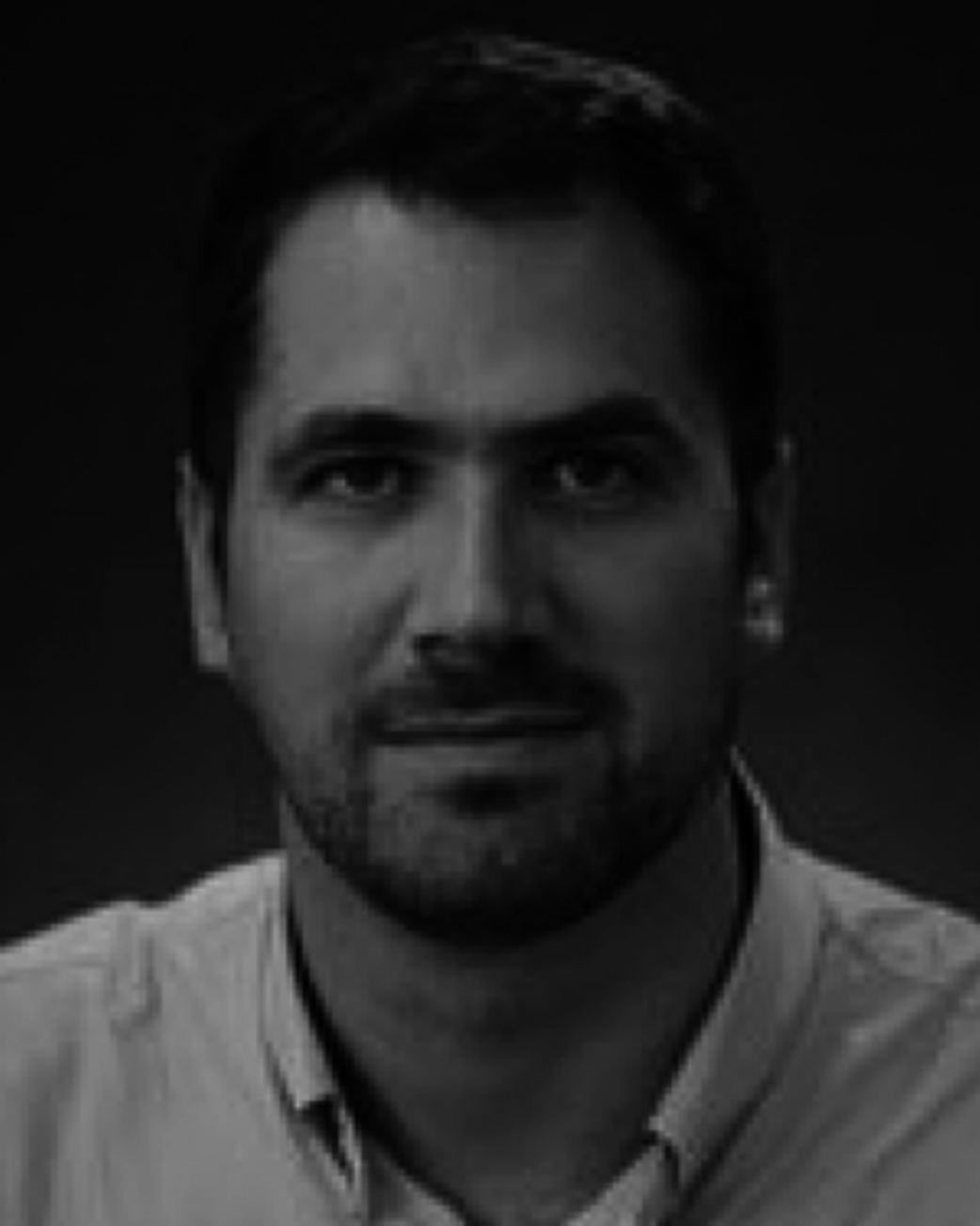}}]{Petros Spachos} received the Diploma in electronic and computer engineering degree from the Technical University of Crete, Chania, Greece, the M.A.Sc. and Ph.D. degrees in electrical and computer engineering from the University of Toronto, Canada. He was a Postdoctoral Researcher with the University of Toronto from September 2014 to July 2015. He is currently an Assistant Professor with the School of Engineering, University of Guelph, Guelph, ON, Canada. His research interests include experimental wireless networking and mobile computing with a focus on wireless sensor networks, smart cities, and the Internet of Things.
\end{IEEEbiography}
\vspace*{-0.5cm}

\begin{IEEEbiography}[{\includegraphics[width=1in,height=1.25in,clip,keepaspectratio]{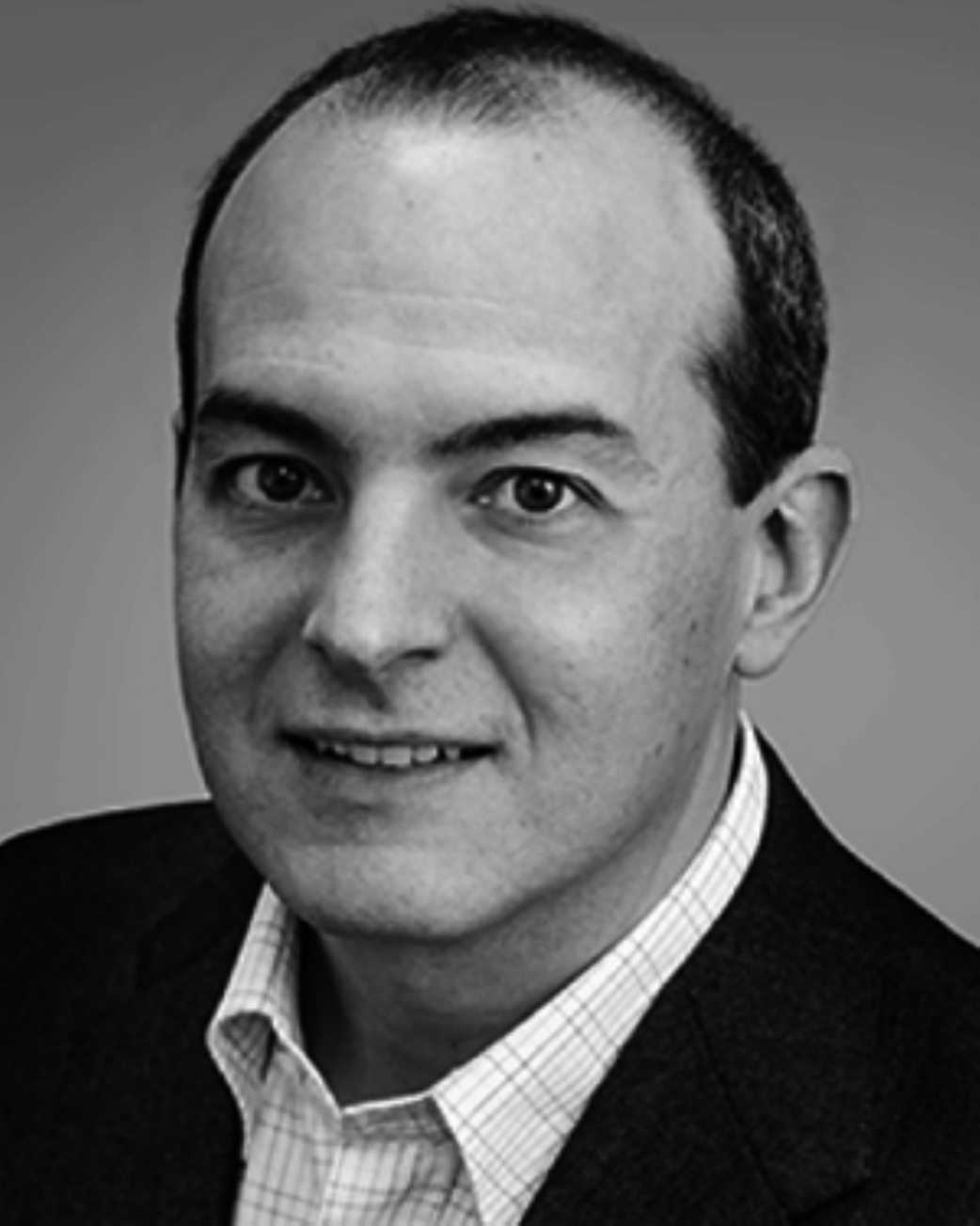}}]
	{Stefano Gregori}
	received the Laurea and Ph.D. degrees in electrical and computer engineering from the University of Pavia, Pavia, Italy. He was with the University of Texas, Dallas, TX, USA, from 2002 to 2004. He is currently a Professor of computer engineering with the University of Guelph, Canada. He is also a Registered Professional Engineer in the Province of Ontario. His research interests include the design, analysis, and characterization of integrated power converters, energy harvesters, and integrated circuits with analog and digital applications.,Dr. Gregori served on the organizing committees for the IEEE Canadian Conference on Electrical and Computer Engineering, and the IEEE International Workshop on Computer Aided Modeling and Design of Communication Links and Networks.
\end{IEEEbiography}
\vspace*{-0.5cm}

\begin{IEEEbiography}[{\includegraphics[width=1in,height=1.25in,clip,keepaspectratio]{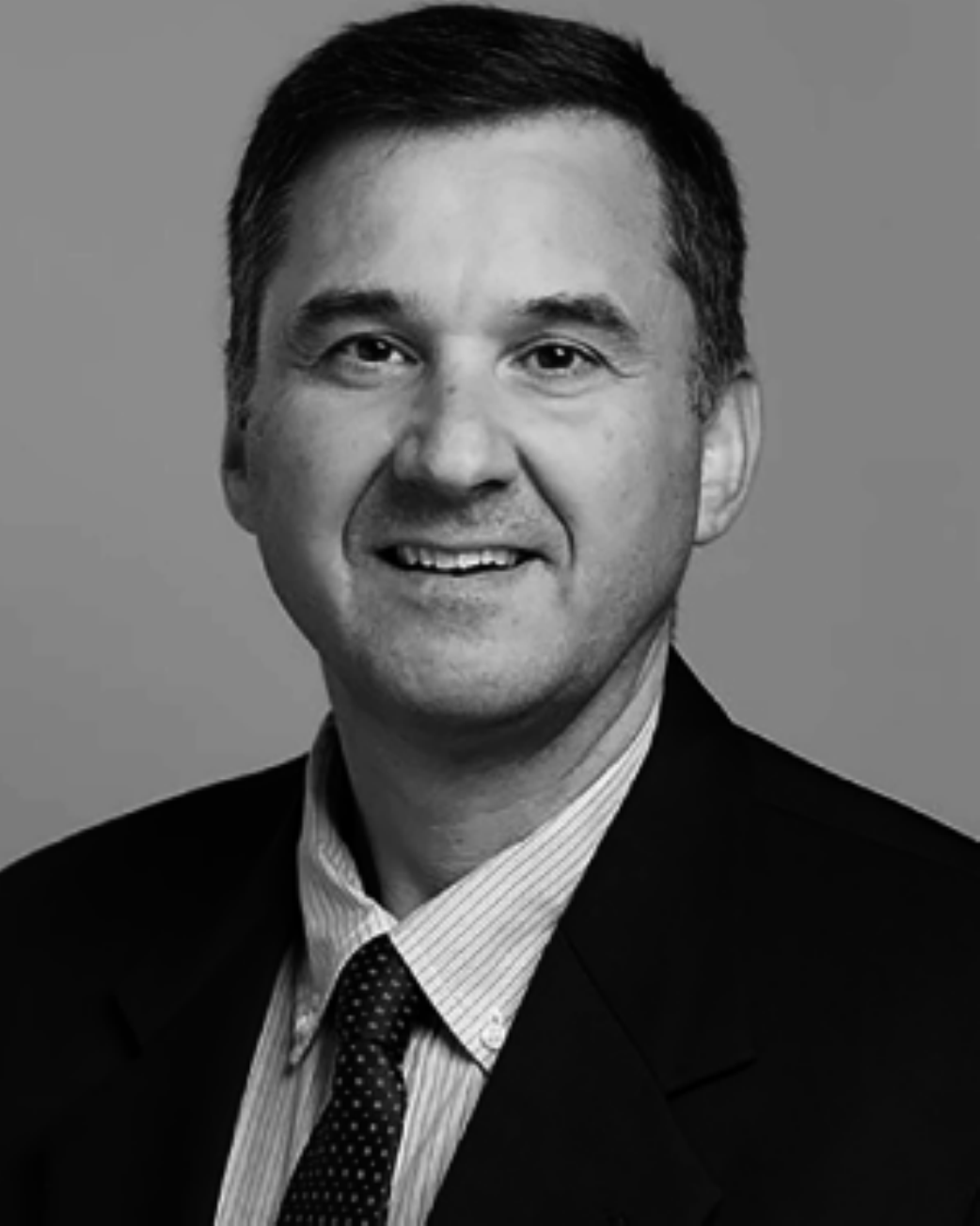}}]{Konstantinos N. Plataniotis} is currently a Professor and the Bell Canada Chair in multimedia with the ECE Department, University of Toronto. He is also the Founder and the Inaugural Director-Research of the Identity, Privacy, and Security Institute (IPSI), University of Toronto. He was the Director of the Knowledge Media Design Institute (KMDI), University of Toronto, from January 2010 to July 2012. He is a Registered Professional Engineer in Ontario. Among his publications in these fields are the recent books WLAN Positioning Systems (2012) and Multilinear Subspace Learning: Reduction of Multidimensional Data (2013). His research interests include knowledge and digital media design, multimedia systems, biometrics, image and signal processing, communications systems, and pattern recognition.  He has served as the Editor-in-Chief for the IEEE Signal Processing Letters.
\end{IEEEbiography}



\end{document}